\documentclass[%
 reprint,
 superscriptaddress,
 amsmath,amssymb,
 aps,
 prl,
 floatfix,
 twocolumn
]{revtex4-2}

\newcommand{\bra}[1]{\langle #1\rvert}
\newcommand{\ket}[1]{\lvert #1\rangle}
\newcommand{\ip}[2]{\bra{#1} #2\rangle}
\newcommand{\op}[2]{\ket{#1} \bra{#2}}
\newcommand{\exv}[1]{\langle #1\rangle}

\DeclareMathOperator{\Tr}{Tr}

\usepackage{graphicx}
\usepackage{dcolumn}
\usepackage{bm}
\usepackage{hyperref}

\usepackage{tikz}
\usetikzlibrary{calc}

\usepackage{amsmath}
\usepackage{upgreek}
\usepackage{physics}
\usepackage{isomath}
\usepackage{placeins}
\usepackage{comment}
\usepackage{mathrsfs}

\begin{document}


\title{Optimal Generators for Quantum Sensing}
\author{Jarrod T. Reilly}
\thanks{These authors contributed equally to this work.}
\affiliation{JILA, NIST, and Department of Physics, University of Colorado, 440 UCB, Boulder, CO 80309, USA}
\author{John Drew Wilson}
\thanks{These authors contributed equally to this work.}
\affiliation{JILA, NIST, and Department of Physics, University of Colorado, 440 UCB, Boulder, CO 80309, USA}
\author{Simon B. J\"ager}
\affiliation{Physics Department and Research Center OPTIMAS, Technische Universit\"at Kaiserslautern, D-67663, Kaiserslautern, Germany}
\author{Christopher Wilson}
\affiliation{Department of Physics, Cornell University, Ithaca, New York 14853, USA}
\author{Murray J. Holland}
\affiliation{JILA, NIST, and Department of Physics, University of Colorado, 440 UCB, Boulder, CO 80309, USA}

\date{\today}

\pacs{Valid PACS appear here}

\begin{abstract}
We propose a computationally efficient method to derive the unitary evolution that a quantum state is most sensitive to.
This allows one to determine the optimal use of an entangled state for quantum sensing, even in complex systems where intuition from canonical squeezing examples breaks down.
In this paper we show that the maximal obtainable sensitivity using a given quantum state is determined by the largest eigenvalue of the quantum Fisher information matrix (QFIM) and the corresponding evolution is uniquely determined by the coinciding eigenvector.
Since we optimize the process of parameter encoding rather than focusing on state preparation protocols, our scheme is relevant for \emph{any} quantum sensor.
This procedure naturally optimizes multiparameter estimation by determining, through the eigenvectors of the QFIM, the maximal set of commuting observables with optimal sensitivity.
\end{abstract}

{
\let\clearpage\relax
\maketitle
}

\emph{Introduction.---}
Advances in quantum sensing technologies including atomic clocks~\cite{Chou,Bothwell}, inertial sensors~\cite{Qvarfort,Richardson,Templier}, gravitational wave detectors~\cite{Aasi,Abbott,Abbott2,Tse}, and biosensors and tissue imaging devices~\cite{Taylor} revolutionize the way we understand the world around us.
The development of sensing devices using parameter estimation is at the core of the ever-growing field of quantum metrology~\cite{Pezze}.
Moreover, quantum sensors can make use of quantum entanglement to surpass the standard quantum limit (SQL) and simultaneously have increased robustness against fluctuations that harm the measurement process~\cite{Teklu,Brivio,Genoni}.
One of the greatest challenges in developing quantum sensors is the generation of metrologically useful entanglement.
Many schemes rely on dynamics in which the quantum state evolution can be intuitively understood.
This provides insight about the final state so it may then be manipulated to utilize its entanglement for a given sensing purpose.
For example, analytic solutions have been developed for one-axis twisting (OAT)~\cite{Kitagawa,Pezze,Pezze2,Agarwal} to track the rotation axis the state is most sensitive to. 

Many theoretical techniques have been developed to determine the metrological usefulness of a state for a given sensing purpose~\cite{Degan,Pezze}.
In particular, the quantum Fisher information (QFI) represents the maximum achievable precision of measuring a specific parameter~\cite{Holevo} and is a sufficient entanglement witness~\cite{Hyllus,Toth}.
However, this assumes a particular evolution and thus fails to shed light on what evolution is optimal when intuition from canonical squeezing examples breaks down.
This is the case in higher dimensional systems where the dynamics cannot be represented on a single collective Bloch sphere~\cite{Chen,Kroeze,Blatt,Lang,Finger,ActuallyJarrodsPaper} and so the potential gain from entanglement cannot be readily determined.

In this Letter, we develop a procedure that finds the physical evolution that a prepared quantum state $\hat{\rho}$ is most sensitive to.
We utilize the quantum Fisher information matrix (QFIM) in which the diagonal elements are the QFI for each single parameter~\cite{Liu,Fiderer}, while the off-diagonal elements represent correlations between two parameters~\cite{Pezze3}.
More fundamentally, the QFIM has a deep connection to distances between quantum states in the language of quantum state geometry~\cite{BengtssonZyczkowski,Campos,Facchi,Provost,Braunstein,Sidhu}.
We use this geometric formalization to show that one can find the optimal evolution by diagonalizing the QFIM.
The largest eigenvalue of the QFIM is the maximum achievable QFI for single parameter estimation and the corresponding eigenvector gives the evolution that achieves this maximum sensitivity. 
Our procedure can also be used for multiparameter estimation with the potential to sense vector or tensorial quantities beyond the SQL~\cite{Templier,Lee,Thiele,Jekeli,Sedlacek,Baumgratz,Kaubruegger}.

To be clear, the purpose of our work is not to propose protocols to create entangled states for quantum sensing. 
Instead, we consider the state fixed and seek to quantify its sensitivity to all possible evolutions, which allows us to determine its full potential for quantum sensing. 
This makes our method useful for \emph{any} preparation scheme of metrologically useful entangled states, assuming the subsequent metrological application is a continuous process.
For practical purposes, this method means one could determine the QFIM of a state, diagonalize it, and then rotate the state until the optimal generator determined here matches the Hamiltonian for a given sensing purpose.
This is a natural consideration because highly entangled states are difficult to engineer while rotations of entangled states are more easily controlled~\cite{Jaksch,Solinas}.

The utility of optimization via QFIM diagonalization becomes clear when one considers the dimensionality of group structures that are often used as the basis for quantum metrological interactions~\cite{Kiselev,Du,Yurke}.
For the case of $\mathrm{SU} (n)$ systems, one has $\dim[\mathfrak{su}(n)] = n^2 - 1$ where $\mathfrak{su}(n)$ is the algebra that generates the group $\mathrm{SU}(n)$ under exponentiation.
To find the optimal generator of evolution, one would have to optimize over the span of $n^2 - 1$ operators which is equivalent to searching an entire $\mathcal{S}^{n^2 - 2}$ hypersphere. 
Instead, the QFIM procedure only requires one to find the eigenvector with the largest eigenvalue of an $(n^2 - 1) \times (n^2 - 1)$ matrix.

\emph{Formalism.---}
We start now by introducing the general formalism. 
To work out a procedure to find the optimal generator, we adopt the language of quantum state geometry [see Supplemental Material (SM)~\cite{suppMat}].
Consider a Hilbert space $\mathcal{H}$ of dimension $d$, with a set of quantum states $\rho(\textbf{x})$. 
Here, the states are parameterized by some ordered list of $n$ coordinates, $\textbf{x} = (x^1,\dots,x^n)$, that are associated with physical parameters. 
The set of $\rho(\textbf{x})$ forms a state-manifold which may be equipped with a Riemannian metric in the form of the QFIM, 
\begin{equation} \label{FisherMetric}
    ds^2 = \mathcal{F}_{\mu \nu} \ dx^\mu dx^\nu,
\end{equation}
with the definitions
\begin{equation}
    \mathcal{F}_{\mu \nu} = \frac{1}{2} \Tr[ \rho \{ \hat{L}_\mu, \hat{L}_\nu \} ], \quad \partial_\mu \rho = \frac{1}{2} \left( \rho \hat{L}_\mu + \hat{L}_\mu \rho \right).
\end{equation}
Here, $\{\hat{A},\hat{B}\} = \hat{A} \hat{B} + \hat{B} \hat{A}$ is the anti-commutator, $\partial_\mu = \partial/\partial x^\mu$, and $\hat{L}_\mu$ is the symmetric logarithmic derivative~\cite{Braunstein} with respect to the coordinate $x^\mu$.

The set of tangent vector fields on the state-manifold represent all potential quantum operations under which the state may evolve.
This is physically equivalent to a set of derivatives, such that any tangent vector may be expanded as $\vec{V} = V^\mu \partial_\mu$. 
From Eq.~\eqref{FisherMetric}, we can then understand the QFI metric $\mathcal{F}_{\mu \nu}$ as the inner product between vectors at the point $\textbf{x}$~\cite{suppMat}: $\langle\vec{V}, \vec{W}\rangle_{\textbf{x}} = \mathcal{F}_{\mu \nu} V^\mu W^\nu$.
In other words, when the QFIM is used to define the interval $ds^2$, it can be intuitively understood as a differential path length across the quantum state space. 

A natural consequence of this interpretation of the QFIM is that the vector whose magnitude is maximized under the QFIM's inner product, labeled $\vec{\mathcal{O}}$, uniquely determines the infinitesimal rotation which changes the state most rapidly.
The magnitude of $\vec{\mathcal{O}}$ is then inversely proportional to the quantum Cr\'amer-Rao bound (QCRB), and thus determines the evolution in which the quantum state is most sensitive to.
Calculating $\vec{\mathcal{O}}$ and its magnitude is equivalent to finding the eigenvector with the largest eigenvalue of $\mathcal{F}_{\mu \nu}$ when treated as a matrix~\cite{suppMat}, 
\begin{equation} \label{QFIMeigenvector} 
    \bm{\mathcal{F}} \vec{\mathcal{O}}^\mu = \lambda_{\mathrm{max}} \vec{\mathcal{O}}^\mu,
\end{equation}
where $\vec{\mathcal{O}}^\mu$ is the column vector representation of $\mathcal{O}^\mu$.

In the case that the parameterization of a state may be described unitarily, $\rho(\textbf{x}) \equiv U(\textbf{x}) \rho(0) U^\dagger(\textbf{x})$, we may further simplify this process since the geometric structure is inherited from the unitary group $\mathrm{U} (\mathcal{H})$~\cite{suppMat,Aniello,Carinena}.
If $\rho(0) = \op{\Psi}$ for some prepared state $\ket{\Psi}$, we can study pure states $U(\textbf{x}) \ket{\Psi}$ belonging to the state-manifold.
Here, the expression for $\mathcal{F}_{\mu \nu}$ simplifies to
\begin{equation}
    \mathcal{F}_{\mu \nu} = 2 \exv{ \{ \hat{G}_\mu , \hat{G}_\nu \} }_{\Psi} - 4 \exv{ \hat{G}_\mu }_{\Psi} \exv{ \hat{G}_\nu }_{\Psi},
\end{equation}
which matches the Fubini-Study metric~\cite{Kolodrubetz}.
We can further understand derivatives at $\textbf{x}$ as
\begin{equation} \label{Derivatives} 
    \partial_\mu U(\textbf{x}) \ket{\Psi} = -i \hat{G}_\mu U(\textbf{x}) \ket{\Psi}, \quad \partial_\mu \equiv -i \hat{G}_{\mu},
\end{equation}
where $-i \hat{G}_\mu \in \mathfrak{u}(\mathcal{H})$ belongs to the Lie algebra. 
Therefore, any vector $\vec{V}$ naturally defines a generator on the Hilbert space according to Eq.~\eqref{Derivatives}, where $V^\mu$ is a set of coefficients associated with the observables $\hat{G}_\mu$ in a Hamiltonian.
The determination of $\vec{\mathcal{O}} = -i \mathcal{O}^\mu \hat{G}_\mu$ is thus equivalent to finding the optimal generator $\hat{\mathcal{G}} = \mathcal{O}^\mu \hat{G}_\mu$.
Here, the QCRB may be artificially lowered by choosing larger coefficients $\tilde{\mathcal{O}}^\mu$ and claiming this leads to a metrological advantage.
As a result, we enforce that $\vec{\mathcal{O}}$ is normalized with respect to the operator basis, $\sum_\mu \left( \mathcal{O}^\mu \right)^2 = 1$.
By further defining a suitable norm $\mathcal{C}$ such that $\Tr[\hat{G}_{\mu}\hat{G}_{\nu}] = \mathcal{C} \delta_{\mu \nu}$~\cite{suppMat}, the SQL is formally defined for $\mathrm{SU}(n)$ systems at the particle number $N$.
This also defines the Heisenberg limit (HL), which is the fundamental sensitivity bound originating from the Heisenberg uncertainty principle~\cite{Holland,Degan}, at $N^2$.

\emph{Squeezing in a SU(2) system.---}
To demonstrate the validity of our QFIM diagonalization procedure, we first consider states created by nonlinear interactions between $N$ two-level particles with an underlying $\mathrm{SU}(2)$ structure.
Each particle's states are labeled with ground state $\ket{d}$ and excited state $\ket{u}$.
We use the Schwinger boson representation~\cite{Schwinger} for two modes with creation operators $\hat{d}^{\dagger}$ and $\hat{u}^{\dagger}$ representing the ``creation'' of a particle in the states $\ket{d}$ and $\ket{u}$, respectively.
As shown in Ref.~\cite{Kitagawa}, squeezing a coherent spin state (CSS), 
\begin{equation}
    \ket{\theta, \phi} = \frac{1}{\sqrt{N!}} \left[ \cos \left( \frac{\theta}{2} \right) \hat{u}^{\dagger} + \sin \left( \frac{\theta}{2} \right) e^{i \phi} \hat{d}^{\dagger} \right]^N \ket{0},
\end{equation}
about a single axis may be accomplished with a nonlinear interaction. 
In particular, the OAT Hamiltonian
\begin{equation} \label{H_OAT}
    \hat{H}_{\mathrm{OAT}} = \hbar \chi \hat{J}_z^2 = \frac{\hbar \chi}{4} \left( \hat{u}^{\dagger} \hat{u} - \hat{d}^{\dagger} \hat{d} \right)^2,
\end{equation}
correlates quantum fluctuations by twisting the northern and southern hemispheres of the collective Bloch sphere in opposite directions, leading to a squeezed state with particle-particle entanglement.
\begin{figure}
    \centerline{\includegraphics[width=\linewidth]{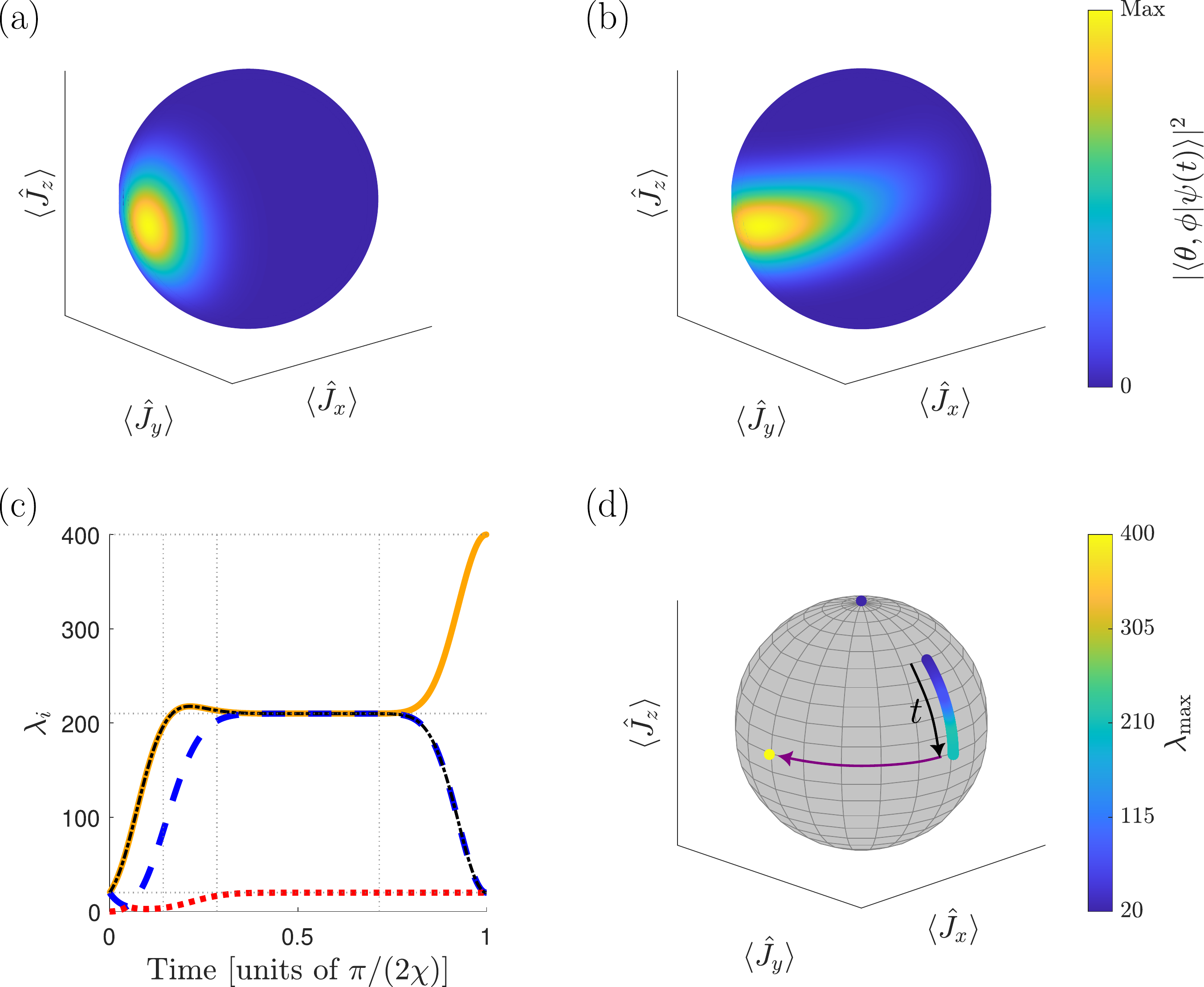}}
    \caption{One-axis twisting with $N = 20$. 
    (a) and (b) Collective Bloch sphere at $t = 0$ and $t = 1 / (\chi N^{\frac{2}{3}})$, respectively. 
    The color represents $\abs{\ip{\theta, \phi}{\psi (t)}}^2$ at each point.
    (c) The three eigenvalues $\lambda_i$ of $\bm{\mathcal{F}}$.
    Also plotted as a black dotted-dashed line is $\mathcal{F}_{\mathrm{OAT}}$ from Eq.~\eqref{F_OAT}.
    (d) Location of the optimal generator during the squeezing process. 
    The color represents the QFI for the given generator.
    The time axis for $t \lesssim \pi / (2 \chi) - 2 / (\chi \sqrt{N})$ is shown with a black arrow, while the discontinuous jump at $t \sim \pi / (2 \chi) - 2 / (\chi \sqrt{N})$ is shown with a purple arrow.
    The eigenvalue then grows at this final axis for the remainder of the process.} 
    \label{fig:OAT}
\end{figure}
We demonstrate squeezing of a CSS initially oriented along $\hat{J}_x = ( \hat{u}^{\dagger} \hat{d} +\hat{d}^{\dagger} \hat{u} ) / 2$, shown in Fig.~\ref{fig:OAT}(a), which reaches an optimally squeezed state at time $t = 1 / (\chi N^{\frac{2}{3}})$, shown in Fig.~\ref{fig:OAT}(b).

We now examine this well-known squeezing example through the lens of QFIM diagonalization. 
The operator basis of the $\mathrm{SU}(2)$ group is the collective operators $\hat{G}_{\mu} \in \{ \hat{J}_x, \hat{J}_y, \hat{J}_z \}$, where $\hat{J}_y = i ( \hat{d}^{\dagger} \hat{u} - \hat{u}^{\dagger} \hat{d}) / 2$.
Therefore, Eq.~\eqref{QFIMeigenvector} requires the diagonalization of a $3 \times 3$ matrix $\bm{\mathcal{F}}$.
Figure~\ref{fig:OAT}(c) shows the three eigenvalues of $\bm{\mathcal{F}}$ during the squeezing process.
At $t = 0$, the eigenvectors $\vec{Y}^\mu = (0, 1, 0)^T$ and $\vec{Z}^\mu = (0, 0, 1)^T$ have degenerate eigenvalues at the SQL, $\lambda_{\mathrm{max}} = N$.
The third eigenvector $\vec{X}^\mu = (1, 0, 0)^T$ has a zero eigenvalue, showing the underlying symmetry of the initial CSS.
The degenerate eigenvalues split as squeezing begins.
As shown in Fig.~\ref{fig:OAT}(c), we find perfect agreement between the largest eigenvalue of $\bm{\mathcal{F}}$ and 
the analytical solution~\cite{Pezze,Pezze2} during the initial squeezing $t \lesssim 1/(\chi \sqrt{N})$,
\begin{equation} \label{F_OAT}
    \mathcal{F}_{\mathrm{OAT}} = N + \frac{N (N - 1)}{4} \left( A + \sqrt{A^2 + B^2} \right),
\end{equation}
with $A = 1 - \cos^{N - 2}(2 \chi t)$ and $B = 4 \sin(\chi t) \cos^{N - 2}(\chi t)$.
We emphasize that this analytical result is found using the exact solution of the squeezing dynamics which allows one to extract the maximum QFI. 
Instead, with the help of the QFIM, we do not require any such insight into the state and yet can still efficiently find the maximum QFI numerically, deriving the eigenvector visible in Fig.~\ref{fig:OAT}(d) displaying the optimal generator $\hat{\mathcal{G}}$.
However, we will see that the QFIM eigendecomposition offers its own insights into symmetries at points of a given system's dynamics.

After $t = 0$, the symmetry of the CSS is broken, and the optimal generator jumps to $\hat{\mathcal{G}} = \sin(\delta) \hat{J}_z + \cos(\delta) \hat{J}_y$, where we find perfect agreement with the expression $\delta = \arctan(B / A) / 2$ given in Ref.~\cite{Kitagawa}. 
As squeezing progresses, the optimal generator then rotates towards the equator.
At $t \sim 2/(\chi \sqrt{N})$, the first two eigenvalues become degenerate with the associated eigenvectors $\vec{X}^{\mu}$ and $\vec{Y}^{\mu}$, once again showing an underlying symmetry of the state~\cite{Agarwal}.
This symmetry is then broken at $t \sim \pi / (2 \chi) - 2/(\chi \sqrt{N})$, causing the two largest eigenvalues to split and a discontinuous jump of the optimal rotation axis from $\vec{Y}^{\mu}$ to $\vec{X}^{\mu}$ [purple arrow in Fig.~\ref{fig:OAT}(d)].
Therefore, Eq.~\eqref{F_OAT} no longer calculates the maximum QFI because it corresponds to rotations about $\hat{J}_y$ when $t \gtrsim 2/(\chi \sqrt{N})$. 
We find that $\mathcal{F}_{\mathrm{OAT}}$ follows the second eigenvalue down to the SQL while the largest eigenvalue grows to the HL, $\lambda_{\mathrm{max}} = N^2$.
The final three eigenvalues, one at HL and two at SQL, are only possible in $\mathrm{SU}(2)$ systems with a NOON state, which matches the analysis of Ref.~\cite{Agarwal}.
Having demonstrated that the well-known results of OAT follow naturally from the diagonalization of the QFIM, we now turn to a higher dimensional system in which analytical results for the maximum QFI and optimal generator cannot readily be obtained. 

\emph{Squeezing in higher dimensonal systems.---}
We consider a $N$-body system in which the constitute particles now have four states $\ket{u}$, $\ket{d}$, $\ket{s}$, and $\ket{c}$.
We again utilize Schwinger bosons with corresponding creation operators $\hat{u}^{\dagger}$, $\hat{d}^{\dagger}$, $\hat{s}^{\dagger}$, and $\hat{c}^{\dagger}$.
Here, the linear dynamics are described by the $\mathrm{SU}(4)$ group with six $\mathfrak{su}(2)$ sub-algebras.
Each sub-algebra has the associated raising operators~\cite{Xu} $\hat{\mathcal{Q}}^+ = \hat{u}^{\dagger} \hat{d}$, $\hat{\Sigma}^+ = \hat{s}^{\dagger} \hat{c}$, $\hat{\mathcal{M}}^+ = \hat{u}^{\dagger} \hat{c}$, $\hat{\mathcal{N}}^+ = \hat{s}^{\dagger} \hat{d}$, $\hat{\mathcal{U}}^+ = \hat{u}^{\dagger} \hat{s}$, and $\hat{\mathcal{V}}^+ = \hat{c}^{\dagger} \hat{d}$.
These operators define the Hermitian components of each algebra according to $\hat{O}_x = (\hat{O}^+ + \hat{O}^-) / 2$, $\hat{O}_y = -i (\hat{O}^+ - \hat{O}^-) / 2$, and $\hat{O}_z = [\hat{O}^+,\hat{O}^-]/2$.
We can then create an operator basis that spans $\mathfrak{su}(4)$ with 15 operators that satisfy the orthonormality property~\cite{suppMat}:
\begin{equation}
    \begin{aligned}
\hat{G}_{\mu} \in \{ & \hat{\mathcal{Q}}_x, \hat{\mathcal{Q}}_y, \hat{\mathcal{Q}}_z, \hat{\Sigma}_x, \hat{\Sigma}_y, \hat{\Sigma}_z, \hat{\mathcal{M}}_x, \hat{\mathcal{M}}_y, \\
& \hat{\mathcal{N}}_x, \hat{\mathcal{N}}_y, \hat{\mathcal{P}}_z, \hat{\mathcal{U}}_x, \hat{\mathcal{U}}_y, \hat{\mathcal{V}}_x, \hat{\mathcal{V}}_y \},
    \end{aligned}
\end{equation}
where $\hat{\mathcal{P}}_z = ( \hat{\mathcal{M}}_z - \hat{\mathcal{N}}_z ) / \sqrt{2}$.

We prepare the state via the nonlinear interaction
\begin{equation} \label{H_TAT}
    \hat{H}_{TAT} = \hbar \chi \left( \hat{\mathcal{Q}}^+ + \hat{\Sigma}^+ \right) \left( \hat{\mathcal{Q}}^- + \hat{\Sigma}^- \right) = 2 \hbar \chi \hat{E}^+ \hat{E}^-,
\end{equation}
which causes twisting about three of the axes of a 15-dimensional collective hypersphere~\cite{ActuallyJarrodsPaper}.
Here, we have introduced three $\mathrm{SU}(2)$ subgroups $\mathfrak{J}$, $\mathfrak{K}$, and $\mathfrak{E}$ generated by algebras with raising operators $\hat{J}^+ = (\hat{\mathcal{M}}^+ + \hat{\mathcal{N}}^+) / \sqrt{2}$, $\hat{K}^+ = (\hat{\mathcal{U}}^+ + \hat{\mathcal{V}}^+) / \sqrt{2}$, and $\hat{E}^+ = (\hat{\mathcal{Q}}^+ + \hat{\Sigma}^+) / \sqrt{2}$, respectively.
The $\mathfrak{J}$ and $\mathfrak{K}$ algebras might represent the dynamics of the internal and external degrees of freedom of atoms in a dispersive Kapitza-Dirac cavity, while the $\mathfrak{E}$ algebras represents the entanglement-generating processes~\cite{ActuallyJarrodsPaper,suppMat}.

\begin{figure}
    \centerline{\includegraphics[width=\linewidth]{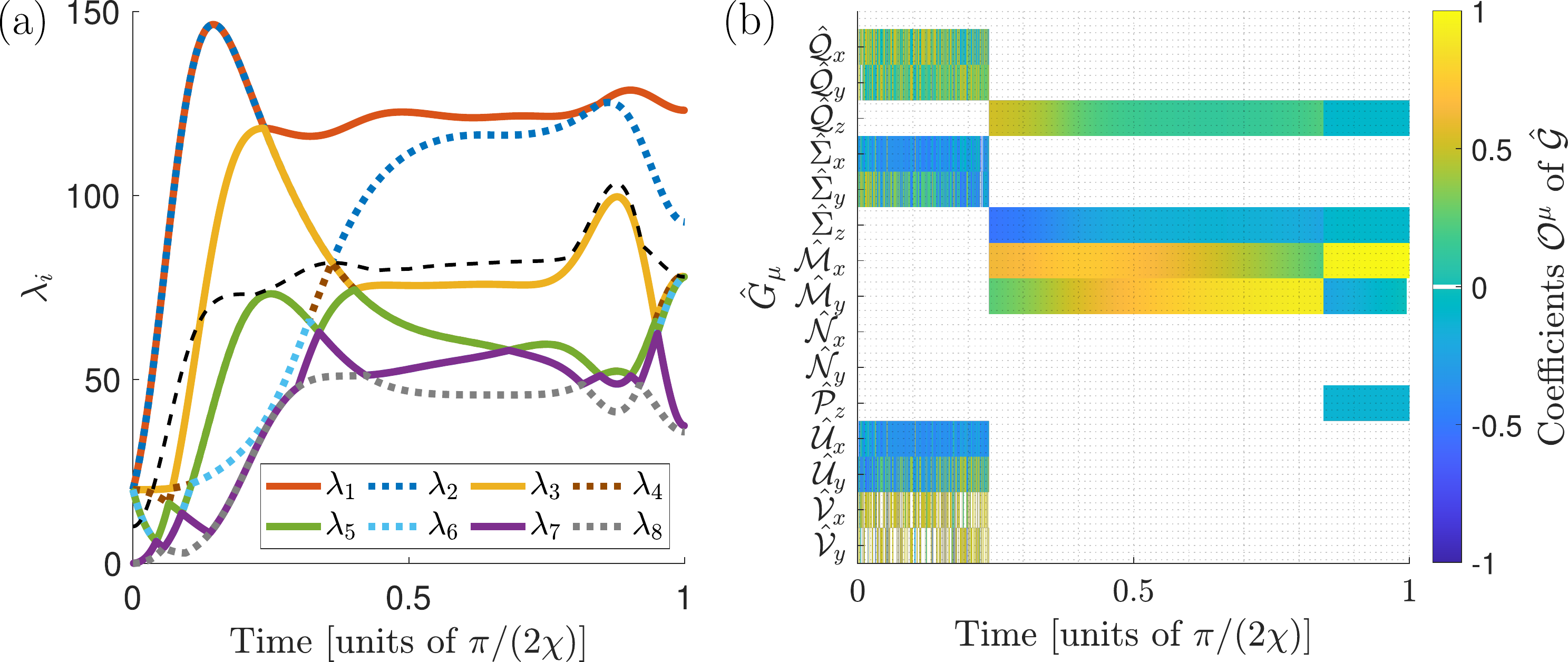}}
    \caption{Three-axis twisting of a $\mathrm{SU}(4)$ system with $N = 20$.
    (a) The largest eight eigenvalues $\lambda_i$ of $\bm{\mathcal{F}}$. 
    Also shown as a black dashed line is the largest QFI from an operator in the $\mathfrak{J}$, $\mathfrak{K}$, and $\mathfrak{E}$ subgroups.
    (b) Coefficients $\mathcal{O}^\mu$ and corresponding basis operator $\hat{G}_\mu$ of the optimal generator $\hat{\mathcal{G}} = \mathcal{O}^\mu \hat{G}_\mu$.}
    \label{fig:TAT}
\end{figure}
When we begin in a simultaneous eigenstate of $\hat{J}_x$ and $\hat{K}_y$, $\ket{\psi_0} = (N!)^{-\frac{1}{2}} \exp[- i \hat{J}_y \pi / \sqrt{2}] (\hat{u}^{\dagger})^N \ket{0}$, the Hamiltonian Eq.~\eqref{H_TAT} causes squeezing as well as non-trivial entanglement between $\mathfrak{J}$ and $\mathfrak{K}$. 
We display the dynamics of the QFIM eigenvalues for $N = 20$ in Fig.~\ref{fig:TAT}(a) and the eigenvector of the QFIM with the largest eigenvalue in Fig.~\ref{fig:TAT}(b).
Figure~\ref{fig:TAT}(a) also displays the maximum QFI from operators in the $\mathfrak{J}$, $\mathfrak{K}$, and $\mathfrak{E}$ subgroups, which were the generators considered in Ref.~\cite{ActuallyJarrodsPaper}. 
At $t = 0$, the largest six eigenvalues are degenerate at the SQL, $\lambda_i = N$, indicating that the starting state is a generalized CSS~\cite{Perelomov}.
This mirrors the symmetry between $\vec{Y}^{\mu}$ and $\vec{Z}^{\mu}$ initially in OAT, but now over three $\mathrm{SU}(2)$ subgroups~\cite{Yukawa}.
As the squeezing begins, the largest two eigenvalues grow until $t \sim 1 / (\chi \sqrt{N})$ where they reach a maximum value of $\lambda_{\mathrm{max}} \sim 146 \approx 0.366 N^2$. 
This degeneracy can be seen in Fig.~\ref{fig:TAT}(b) as the optimal generator jumps back and forth between two operators for the beginning of the squeezing process. 
These two degenerate eigenvalues subsequently fall until they cross the third largest eigenvalue at $t \sim 5 / (3 \chi \sqrt{N})$, corresponding to a discontinuous jump in Fig.~\ref{fig:TAT}(b).
The eigenvalue corresponding to a rotation axis close to $\hat{\mathcal{M}}_x$ then grows rapidly, eventually becoming the largest eigenvalue at $t \sim \pi / (2 \chi) - 1 / (\chi \sqrt{N})$.
This analysis highlights that the QFIM diagonalization unravels the complicated nonlinear dynamics of the high dimensional quantum system.
In fact, with its help, we find that at all times the state has a higher sensitivity than what was shown in Ref.~\cite{ActuallyJarrodsPaper}. 

\emph{Multiparameter estimation.---}
So far, we have focused on optimizing single parameter estimation. 
However, our QFIM diagonalization scheme inherently optimizes multiparameter estimation as well by finding multiple eigenvectors of the QFIM whose complimentary generators commute with one another.
This, in turn,  could be used in quantum sensors that aim to infer multiple parameters beyond the SQL simultaneously.
As an example, at $t = \pi / (4 \chi)$ in Fig.~\ref{fig:TAT}(a), the generators associated with the eigenvalues $\lambda_1 = 0.307 \, N^2$, $\lambda_3 = 0.189 \, N^2$, and $\lambda_8 = 0.117 \, N^2$ all commute with one another, meaning one could carry out simultaneous estimation beyond the SQL for all three of the corresponding parameters.
The associated generators go as $\hat{\mathcal{G}}_1 = c_1 \sqrt{2} \hat{K}_z + c_2 (\hat{\mathcal{M}}_x + \hat{\mathcal{M}}_y)$, $\hat{\mathcal{G}}_3 = c_2 \sqrt{2} \hat{K}_z - c_1 (\hat{\mathcal{M}}_x + \hat{\mathcal{M}}_y)$, and $\hat{\mathcal{G}}_8 = c_3 \hat{\mathcal{N}}_x + c_4 \hat{\mathcal{N}}_y$, with real coefficients $c_i$ that satisfy the normalization condition.
For the case of the spin-momentum $\textrm{SU}(4)$ system considered in Ref.~\cite{ActuallyJarrodsPaper}, a portion of these generators may be found to correspond to interactions which are more physically accessible than the whole generator is [see SM~\cite{suppMat} for details].
In this physical example, $\hat{K}_z$ could correspond to a linear acceleration while $\hat{\mathcal{M}}_x + \hat{\mathcal{M}}_y$ and $c_3 \hat{\mathcal{N}}_x + c_4 \hat{\mathcal{N}}_y$ may correspond to spatially dependent rotations, thereby creating the opportunity for many combinations of useful interferometry~\cite{Lee,Hardman,Malia}.
We thus consider $(\hat{\mathcal{M}}_x + \hat{\mathcal{M}}_y) / \sqrt{2}$, $\hat{K}_z$, and $\hat{\mathcal{G}}_8$ which still have QFIs of $0.300 \, N^2$, $0.195 \, N^2$, and $0.117 \, N^2$, respectively. 
Since these operators are in three commuting sub-algebras, they can be independently rotated to any arbitrary operator in the respective sub-algebra in order to be made relevant for sensing vector quantities or network node interferometry~\cite{Malia,Zhang2,Pelayo}.

More generally, within $\mathrm{SU}(n)$, one is guaranteed sets of $n - 1$ commuting generators~\cite{suppMat,Montgomery,Ku}, thereby guaranteeing sets of $n - 1$ eigenvectors of the QFIM which correspond to simultaneously commuting generators.
One could thus select the eigenvector with the largest eigenvalue and search the remaining eigenvectors to find the set of $n - 1$ generators which mutually commute and have suitable eigenvalues that scale beyond the SQL.
Furthermore, the associated symmetric logarithmic derivatives $\hat{L}_{\mu}$ are guaranteed to commute such that the optimal measurement basis is the same for each parameter.
This ensures that the QCRB is always simultaneously attainable for all $n - 1$ parameters as the elements of the Uhlmann curvature matrix $\mathbb{U}_{\mu \nu} = - i \Tr[\rho [\hat{L}_{\mu}, \hat{L}_{\nu}] ] / 2$ will vanish~\cite{Liu,Carollo,Candeloro,Asjad}.

\emph{Conclusion and outlook.---}
We have demonstrated that the optimal generator for quantum sensing is given by the eigenvector associated with the largest eigenvalue of the QFIM. 
This is a consequence of maximizing differential path lengths through quantum state space when the QFIM is viewed as a Riemannian metric, generalizing the work of Ref.~\cite{Hyllus2} to any metrological process with an underlying Lie group structure.
For the examples we considered, unitary parameterization was assumed, but future steps include examining a channel or hybrid parameterization scheme~\cite{Liu,Nair,Jonsson} using QFIM diagonalization.
Furthermore, our examples have used pure states, but the procedure is equally valid with mixed states and the properly defined tangent vectors.
Here, one must utilize the more general definition of the QFIM given in Ref.~\cite{Liu}.
The use of mixed states is then relevant to experiments where a small amount of entanglement entropy between the system and a bath can be generated through either known or unknown dissipative processes. 

The examples we considered had underlying $\mathrm{SU}(2)$ and $\mathrm{SU}(4)$ group dynamics.
Already in the case of $\mathrm{SU}(4)$, one finds that more care must be taken compared to the $\mathrm{SU}(2)$ case when considering larger group structures.
For one, unitarily rotating the optimal generator to an arbitrary operator is not always possible in larger group structures~\cite{suppMat}.
We also outline in the SM~\cite{suppMat} how to extend our work to general $\mathrm{SU}(n)$ systems with an algorithm to generate an orthogonal operator basis that spans the quantum state space.
Moreover, the underlying formalism of this Letter extends to any dynamical group structure.
This makes our procedure relevant to systems described by $\mathrm{Sp}(n,\mathbb{R})$~\cite{Wunsche,Howard}, $\mathrm{SU}(m,n)$~\cite{Caves}, or translational groups~\cite{Sainadh}, for example.

Interestingly, there have been recent efforts to experimentally infer the quantum geometric tensor~\cite{Ozawa,Yu,Yu2,Zhang}, which is related to the QFI metric through its real component~\cite{suppMat,Kolodrubetz}.
This leads to the prospect of finding the optimal generator for quantum sensing without the need for a full theoretical model, only an understanding of the underlying symmetries.
This is necessary for complex systems where such models are difficult to derive or fully simulate.
Our QFIM diagonalization procedure thus opens an exciting avenue for experiments with complex systems~\cite{Lang,Bishof,Song,Sonderhouse,Taie,Gorshkov,Blatt,Finger,Luo}, whose current interest is not parameter estimation, to naturally test if the experiment can be useful as a quantum sensor and how to use any generated entanglement in an efficient manner. 
In addition, we can combine numerical approaches with QFIM diagonalization for these complex systems, which is relevant for quantum optical control and machine learning methods that have been used effectively for quantum design tasks~\cite{Liu2,Liu3,Basilewitsch,Shushen,Chih,Chih2,LeDesma}.

\begin{acknowledgments}
We thank John Cooper, Klaus M\o{}lmer, and Joshua Combes for useful discussions.
This research was supported by
NSF PHY 1734006; NSF OMA 2016244; NSF PHY Grant No. 2207963; and NSF 2231377.
S.B.J. acknowledges support from the Deutsche Forschungsgemeinschaft (DFG): Projects A4 and A5 in SFB/Transregio 185: “OSCAR”.

J.T.R. and J.D.W. contributed equally to this work.
\end{acknowledgments}

\bibliography{references.bib}

\appendix

\end{document}



\title{Supplemental Material: Optimal Generators for Quantum Sensing}
\author{Jarrod T. Reilly}
\thanks{These authors contributed equally to this work.}
\affiliation{JILA, NIST, and Department of Physics, University of Colorado, 440 UCB, Boulder, CO 80309, USA}
\author{John Drew Wilson}
\thanks{These authors contributed equally to this work.}
\affiliation{JILA, NIST, and Department of Physics, University of Colorado, 440 UCB, Boulder, CO 80309, USA}
\author{Simon B. J\"ager}
\affiliation{Physics Department and Research Center OPTIMAS, Technische Universit\"at Kaiserslautern, D-67663, Kaiserslautern, Germany}
\author{Christopher Wilson}
\affiliation{Department of Physics, Cornell University, Ithaca, New York 14853, USA}
\author{Murray J. Holland}
\affiliation{JILA, NIST, and Department of Physics, University of Colorado, 440 UCB, Boulder, CO 80309, USA}

\date{\today}

\pacs{Valid PACS appear here}

\maketitle

\tableofcontents

\section{Operator Basis for Symmetric \texorpdfstring{$\mathrm{SU}(n)$}{SU(n)} Systems} \label{OperatorBasis}
We now outline how to form the operator basis for general $\mathrm{SU} (n)$ systems.
We want our basis operators to be linearly independent $\Tr[\hat{G}_\mu \hat{G}_\nu] = \mathcal{C} \delta_{\mu \nu}$, traceless $\Tr[\hat{G}_\mu] = 0$, and all have the same norm $\mathcal{C}$ defined in Eq.~\eqref{eq:KillingNorm}.
The norm condition ensures the basis operators have the same spectrum $\sigma(\hat{G}_\mu) = \sigma(\hat{G}_\nu)$, so that they can be connected unitarily without need for rescaling.

Since we are considering states which are symmeterized over the constituent particles, we may utilize the Schwinger boson representation~\cite{Schwinger}.
For $n$ states $\ket{a_1}, \ket{a_2}, \ket{a_3}, \ldots, \ket{a_n}$, we define the annihilation operators $\hat{a}_1, \hat{a}_2, \hat{a}_3, \ldots, \hat{a}_n$ which can then form the collective operators
\begin{equation}
    \begin{aligned} 
\hat{\mathcal{A}}_{12}^+ &= \hat{a}_1^{\dagger} \hat{a}_2, \hat{\mathcal{A}}_{13}^+ = \hat{a}_1^{\dagger} \hat{a}_3, \ldots, \hat{\mathcal{A}}_{1n}^+ = \hat{a}_1^{\dagger} \hat{a}_n, \\
\hat{\mathcal{A}}_{23}^+ &= \hat{a}_2^{\dagger} \hat{a}_3, \hat{\mathcal{A}}_{24}^+ = \hat{a}_2^{\dagger} \hat{a}_4, \ldots, \hat{\mathcal{A}}_{(n-1) n}^+ = \hat{a}_{n-1}^{\dagger} \hat{a}_n.
    \end{aligned}
\end{equation}
Defining $\hat{\mathcal{A}}_{ij}^- = (\hat{\mathcal{A}}_{ij}^+)^{\dagger}$, we can find $\binom{n}{2} = n!/[2! (n-2)!] = n(n-1) / 2$ closed $\mathfrak{su} (2)$ sub-algebras by the relations
\begin{equation}
    \hat{\mathcal{A}}_{ij}^x = \frac{\hat{\mathcal{A}}_{ij}^+ + \hat{\mathcal{A}}_{ij}^-}{2}, \; \hat{\mathcal{A}}_{ij}^y = \frac{\hat{\mathcal{A}}_{ij}^+ - \hat{\mathcal{A}}_{ij}^-}{2 i}, \; \hat{\mathcal{A}}_{ij}^z = \frac{\left[ \hat{\mathcal{A}}_{ij}^+, \hat{\mathcal{A}}_{ij}^- \right]}{2}.
\end{equation}
We have, by definition, that the $x$ and $y$ operators from the different sub-algebras are linearly independent, leading to $2 \binom{n}{2} = n(n-1)$ basis operators.
This leaves $n^2 - 1 - 2 \binom{n}{2} = n - 1$ basis operators to be made up from the $\hat{\mathcal{A}}_{ij}^z$ operators.
However, one can clearly see that more care must be taken for the $\hat{\mathcal{A}}_{ij}^z$ operators when they are rewritten as
\begin{equation}
    \hat{\mathcal{A}}_{ij}^z = \frac{\hat{a}_i^{\dagger} \hat{a}_i - \hat{a}_j^{\dagger} \hat{a}_j}{2},
\end{equation}
since the number operator $\hat{a}_i^{\dagger} \hat{a}_i$ appears in multiple $z$ operators.
Thus, these operators will not be linearly independent in general, with $\Tr[\hat{\mathcal{A}}_{ij}^z \hat{\mathcal{A}}_{ik}^z] \neq 0$ as an example. 

To choose the remaining $n - 1$ basis operators, one can first select $n / 2$ linearly independent $\hat{\mathcal{A}}_{ij}^z$ operators if $n$ is even or $(n-1) / 2$ operators if $n$ is odd:
\begin{equation}
    \begin{aligned}
\hat{\mathcal{A}}_{12}^z, \hat{\mathcal{A}}_{34}^z, \ldots, \hat{\mathcal{A}}_{(n-1)n}^z, \quad &n \in 2 \mathbb{N}, \\
\hat{\mathcal{A}}_{12}^z, \hat{\mathcal{A}}_{34}^z, \ldots, \hat{\mathcal{A}}_{(n-2)(n-1)}^z, \quad &n \in 2 \mathbb{N} - 1,
    \end{aligned}
\end{equation}
where $\mathbb{N}$ represents the natural numbers.
Then, one makes normalized operators from alternating indices:
\begin{equation}
    \begin{aligned}
& \mathcal{N}_2 \left[ \hat{\mathcal{A}}_{13}^z + \hat{\mathcal{A}}_{24}^z \right], \mathcal{N}_4 \left[ \hat{\mathcal{A}}_{15}^z + \hat{\mathcal{A}}_{26}^z + \hat{\mathcal{A}}_{35}^z + \hat{\mathcal{A}}_{46}^z \right], \ldots, \\
& \mathcal{N}_{n - 2} \left[ \sum_{i=1}^{\frac{n}{2}-1} \hat{\mathcal{A}}_{(2i - 1)(n-1)}^z + \sum_{j=1}^{\frac{n}{2}-1} \hat{\mathcal{A}}_{(2j)n}^z \right], \quad n \in 2 \mathbb{N}, \\
& \mathcal{N}_2 \left[ \hat{\mathcal{A}}_{13}^z + \hat{\mathcal{A}}_{24}^z \right], \mathcal{N}_4 \left[ \hat{\mathcal{A}}_{15}^z + \hat{\mathcal{A}}_{26}^z + \hat{\mathcal{A}}_{35}^z + \hat{\mathcal{A}}_{46}^z \right], \ldots, \\
& \mathcal{N}_{n - 3} \left[ \sum_{i=1}^{\frac{n-3}{2}} \hat{\mathcal{A}}_{(2i - 1)(n-2)}^z + \sum_{j=1}^{\frac{n-3}{2}} \hat{\mathcal{A}}_{(2j)(n-1)}^z \right], \quad n \in 2 \mathbb{N} - 1,
    \end{aligned}
\end{equation}
where the normalization for operator $\hat{O}$ is given by $\mathcal{N}_i = \sqrt{\mathcal{C} / \Tr[\hat{O}^2]} = \sqrt{\Tr[(\hat{\mathcal{A}}_{12}^z)^2] / \Tr[\hat{O}^2]}$.
This gives another $n/2 - 1$ basis operators for even $n$ and $(n-1)/2 - 1$ more operators for odd $n$.
The final linearly independent basis operator for odd $n$ is then given by
\begin{equation}
    \mathcal{N}_{n - 1} \left[ \hat{\mathcal{A}}_{1n}^z + \hat{\mathcal{A}}_{2n}^z + \ldots + \hat{\mathcal{A}}_{(n-1)n}^z \right], \quad n \in 2 \mathbb{N} - 1.
\end{equation}

\section{Unitary Connection Between Operators}
In this section, we outline a method to find the unitary rotation that takes the optimal generator $\hat{\mathcal{G}}$ to one of physical interest $\hat{\mathcal{Z}}$ in order to sense some specific parameter.
In the case of $\mathrm{SU}(2)$, one can always find the unitary connection between these operators $\hat{\mathcal{Z}} = \hat{U}_c^{\dagger} \hat{\mathcal{G}} \hat{U}_c$.
However, when larger group structures are considered, there is no such guarantee. 
A unitary connection can be found if and only if the operators have the same spectrum $\sigma(\hat{\mathcal{G}}) = \sigma(\hat{\mathcal{Z}})$.
In this case, one can find the operator, $\hat{\mathcal{R}}$, that generates $\hat{U}_c = \exp[ - i \hat{\mathcal{R}} \pi / 2 ]$ by solving a special case of Sylvester's equation~\cite{Bhatia}: 
\begin{equation} \label{SylvestersEq}
    i [ \hat{\mathcal{R}}, \hat{\mathcal{G}} ] = \hat{\mathcal{Z}}.
\end{equation}
A non-unitary connection can be found when $\sigma(\hat{\mathcal{G}}) \neq \sigma(\hat{\mathcal{Z}})$, and there may be some cases where there is experimental relevancy to such solutions~\cite{Torosov,Reyes,Gunther,Hurst,Ibanez}, but we do not pursue these connections here.

To solve Eq.~\eqref{SylvestersEq}, one vectorizes $\hat{\mathcal{R}}$ and $\hat{\mathcal{Z}}$ via the mapping 
\begin{equation} \label{OpMap}
    \hat{O} = \sum_{a,b} o_{a,b} \op{a}{b} \longleftrightarrow \lvert \hat{O} \rangle \rangle = \sum_{a,b} o_{a,b} \ket{a} \otimes \ket{b},
\end{equation}
and then solves
\begin{equation}
    i \left[ \hat{\mathbb{I}} \otimes \hat{\mathcal{G}}^T - \hat{\mathcal{G}} \otimes \hat{\mathbb{I}} \right] \lvert \hat{\mathcal{R}} \rangle \rangle \equiv i \hathat{\mathbb{G}} \lvert \hat{\mathcal{R}} \rangle \rangle = \lvert \hat{\mathcal{Z}} \rangle \rangle,
\end{equation}
with identity operator $\hat{\mathbb{I}}$.
The superoperator $\hathat{\mathbb{G}}$ can not be inverted as $\abs{\hat{\mathbb{I}} \otimes \hat{\mathcal{G}}^T} = \abs{\hat{\mathcal{G}} \otimes \hat{\mathbb{I}}}$. 
However, there are techniques to solve this linear set of equations numerically.
For example, one can sometimes utilize the Moore-Penrose pseudoinverse $\hat{A}^+$~\cite{Penrose,Barata} which satisfies the conditions
\begin{equation}
    \begin{aligned} 
\hat{A} \hat{A}^+ \hat{A} = \hat{A},& \quad \hat{A}^+ \hat{A} \hat{A}^+ = \hat{A}^+ \\
\left( \hat{A} \hat{A}^+ \right)^{\dagger} = \hat{A} \hat{A}^+,& \quad \left( \hat{A}^+ \hat{A} \right)^{\dagger} = \hat{A}^+ \hat{A}.
    \end{aligned}
\end{equation}
Note that $\hat{A}^+ = \hat{A}^{-1}$ when $\hat{A}$ is invertible.
The pseudoinverse exists and is unique for any matrix, but it does not always have a simple algebraic form.

In the case that $\hathat{\mathbb{G}}$ has linearly independent columns, the pseudoinverse becomes
\begin{equation}
    \hathat{\mathbb{G}}^+ = (\hathat{\mathbb{G}}^{\dagger} \hathat{\mathbb{G}})^{-1} \hathat{\mathbb{G}}^{\dagger},
\end{equation}
and corresponds to a left inverse $\hathat{\mathbb{G}}^+ \hathat{\mathbb{G}} = \hat{\mathbb{I}} \otimes \hat{\mathbb{I}}$, such that
\begin{equation}
    \lvert \hat{\mathcal{R}} \rangle \rangle = -i \left( \hathat{\mathbb{G}}^{\dagger} \hathat{\mathbb{G}} \right)^{-1} \hathat{\mathbb{G}}^{\dagger} \lvert \hat{\mathcal{Z}} \rangle \rangle.
\end{equation}
One can then map this back into the original space via Eq.~\eqref{OpMap} and then retrieve the coefficients $\mathcal{R}^\mu$ that compose $\hat{\mathcal{R}} = \mathcal{R}^\mu \hat{G}_\mu$ in order to determine the physical implementation of $\hat{\mathcal{R}}$.
In the more general case, the pseudoinverse can be calculated using the single value decomposition of $\hathat{\mathbb{G}}$~\cite{Barata}.

\section{Example Realization and Additional Analysis of the \texorpdfstring{$\mathrm{SU}(4)$}{SU(4)} System} \label{SU4map}
We now briefly describe an example realization of the $\mathrm{SU}(4)$ Hamiltonian from Eq.~(10) of the Main Text. 
This theoretical model was first presented in Ref.~\cite{ActuallyJarrodsPaper}.

\subsection{Example Realization of the Hamiltonian}
We consider a packet of $N$ two-level atoms with uniform velocity that pass through a dispersive cavity.
The atoms can be prepared in the $\ket{\hbar k / 2}$ state utilizing Kapitza-Dirac pulses~\cite{Li}, and assuming $\hbar N g^2 / (4 \abs{\Delta}) \ll (\hbar k)^2 / m$, they will remain in the momentum subspace spanned by $\ket{\pm \hbar k / 2}$.
Here, $k$ is the wavenumber of the cavity, $g$ is the cavity coupling strength, $\Delta$ is the atom-cavity detuning, and $m$ is the atomic mass.
After the cavity is adiabatically eliminated based on large detuning $\abs{\Delta} \gg \sqrt{N} g$, the Hamiltonian is given by~\cite{ActuallyJarrodsPaper}
\begin{equation} \label{H_full}
    \hat{H} = \hbar \chi \sum_{i,j = 1}^N \hat{s}_i^x \hat{s}_j^x \hat{\sigma}_i^+ \hat{\sigma}_j^-,
\end{equation}
where $\chi = g^2 / (4 \Delta)$, $\hat{\sigma}_i^+ = \op{e}{g}_i$, and $\hat{s}_i^x = (\op{\hbar k / 2}{-\hbar k / 2}_i + \op{-\hbar k / 2}{\hbar k / 2}_i) / 2$.
This Hamiltonian can generate both particle-particle and spin-motion entanglement when the atoms are prepared in the state $\ket{\psi_0} = 2^{-\frac{N}{2}} (\ket{g} + \ket{e})^{\otimes N} \otimes \ket{\hbar k / 2}^{\otimes N}$, where $\ket{g}$ and $\ket{e}$ are the atoms' internal states.

We note the system has an underlying $\mathrm{SU}(4)$ structure and so, in analogy to Ref.~\cite{Xu}, we label
\begin{equation}
    \begin{aligned} 
\ket{u} &\equiv \ket{e, \hbar k / 2}, \quad \ket{d} \equiv \ket{g, -\hbar k / 2}, \\ 
\ket{s} &\equiv \ket{e, -\hbar k / 2}, \quad \ket{c} \equiv \ket{g, \hbar k / 2}.
    \end{aligned}
\end{equation}
If one begins in the symmetric subspace of $\mathrm{SU}(4)$, 
\begin{equation}
    \ket{P_{\alpha, \beta, \gamma, \delta}} = \sqrt{\frac{\alpha! \beta! \gamma! \delta!}{N!}} \mathcal{S} (\ket{u}^{\otimes \alpha} \ket{d}^{\otimes \beta} \ket{s}^{\otimes \gamma} \ket{c}^{\otimes \delta}),
\end{equation}
for $\alpha + \beta + \gamma + \delta = N$ and symmetrizer $\mathcal{S}$, then the system will remain in this subspace.
The subspace scales as $(N+1)(N+2)(N+3)/6 \sim \mathcal{O}(N^3)$ making exact numerical calculations feasible.
Using the associated Schwinger boson representation, Eq.~\eqref{H_full} can be rewritten as 
\begin{equation}
    \hat{H} = \hbar \chi (\hat{u}^{\dagger} \hat{d} + \hat{s}^{\dagger} \hat{c}) (\hat{u} \hat{d}^{\dagger} + \hat{s} \hat{c}^{\dagger}).
\end{equation}
We can now introduce three subgroups $\mathfrak{J}$, $\mathfrak{K}$, and $\mathfrak{E}$, with corresponding algebras $\{ \hat{J}^{\pm}, \hat{J}_z \}$, $\{ \hat{K}^{\pm}, \hat{K}_z \}$, and $\{ \hat{E}^{\pm}, \hat{J}_z \}$, respectively.
These subgroups represent the spin and momentum dipoles and entanglement generating processes between the two subsystems.
Here, we have defined the collective operators 
\begin{equation} \label{CollectiveOps}
    \begin{aligned}
\hat{J}^{\pm} &= \frac{1}{\sqrt{2}} \sum_{j = 1}^N \hat{\sigma}_j^{\pm}, \quad \hat{J}_z = \frac{1}{2 \sqrt{2}} \sum_{j = 1}^N \hat{\sigma}_j^z, \\
\hat{K}^{\pm} &= \frac{1}{\sqrt{2}} \sum_{j = 1}^N \hat{s}_j^{\pm}, \quad \hat{K}_z = \frac{1}{2 \sqrt{2}} \sum_{j = 1}^N \hat{s}_j^z, \\
\hat{E}^{\pm} &= \frac{1}{\sqrt{2}} \sum_{j = 1}^N \hat{\sigma}_j^{\pm} \hat{s}_j^x,
    \end{aligned}
\end{equation}
and noted $\hat{E}_z = [\hat{E}^+, \hat{E}^-] / 2 = \hat{J}_z$.
Writing $\hat{E}^+ = (\hat{u}^{\dagger} \hat{d} + \hat{s}^{\dagger} \hat{c}) / \sqrt{2}$, the Hamiltonian becomes
\begin{equation}
    \hat{H} = 2 \hbar \chi \hat{E}^+ \hat{E}^- = 2 \hbar \chi \left( \hat{E}^2 - \hat{J}_z^2 + \hat{J}_z \right),
\end{equation}
where $\hat{E}^2 = \hat{E}_x^2 + \hat{E}_y^2 + \hat{J}_z^2$.
Notice that the factor of $2$ difference with respect to the Hamiltonian in Ref.~\cite{ActuallyJarrodsPaper} is because of the $\sqrt{2}$ in Eq.~\eqref{CollectiveOps}. 
This factor is included so that the normalization conditions $\sum_\mu (\mathcal{O}^\mu)^2 = 1$ and Eq.~\eqref{eq:KillingNorm} are met with respect to the operator basis defined in Eq.~(9) of the Main Text.

\subsection{Example Realization of the Basis Operators}
Once one diagonalizes the QFIM with respect to the operator basis from Eq.~(9) of the Main Text, one may wish to rotate the state to make certain operators the optimal generators for single or multiparameter estimation.
We therefore discuss how one could realize such rotations in the experimental setup discussed in the previous subsection.
We assume the atomic beam passes through the cavity to become entangled, and then one performs metrology once the atoms leave the cavity.
The atoms will spatially separate into two packets due to their opposite momentum with respect to the cavity axis, which we designate as the $x$ axis, in a similar fashion to recent network node interferometry experiments~\cite{Zhang2,Malia}. 

We begin with rotations on the $\mathcal{M}$ and $\mathcal{N}$ subalgebras, which have the raising operators $\hat{\mathcal{M}}^+ = \hat{u}^{\dagger} \hat{c}$ and $\hat{\mathcal{N}}^+ = \hat{s}^{\dagger} \hat{d}$.
Once the two packets are spatially separated, one can address them individually.
Therefore, these rotations can be achieved by addressing the $\hbar k / 2$ packet for $\mathcal{M}$ and the $- \hbar k / 2$ packet for $\mathcal{N}$.
The lasers that perform these rotations should be oriented in an orthogonal direction from the $x$ axis, which we designate as the $z$ axis.
This ensures that these lasers can perform rotations on the internal states without changing the atoms' $x$-momentum.

We next discuss $\mathcal{Q}$ and $\Sigma$ rotations, generated by the raising operators $\hat{\mathcal{Q}}^+ = \hat{u}^{\dagger} \hat{d}$ and $\hat{\Sigma}^+ = \hat{s}^{\dagger} \hat{c}$.
These can be accomplished by laser pulses oriented in the $x$-direction.
Pulses in the $+x$-direction will cause $\mathcal{Q}$ rotations while pulses in the $-x$-directions drives $\Sigma$ rotations.
One must take care that the laser Rabi frequency $\Omega$ for these pulses is small enough that one does not drive excitations to higher motional states, $\abs{\Omega} \ll \hbar k^2 / m$.

Lastly, we turn to $\mathcal{U}$ and $\mathcal{V}$ rotations, generated by the raising operators $\hat{\mathcal{U}}^+ = \hat{u}^{\dagger} \hat{s}$, and $\hat{\mathcal{V}}^+ = \hat{c}^{\dagger} \hat{d}$.
These rotations can be accomplished using Bragg pulses.
Consider an auxiliary internal excited state $\ket{f}$ from which one can independently drive the $\ket{e} \leftrightarrow \ket{f}$ and $\ket{g} \leftrightarrow \ket{f}$ transitions using different laser wavelengths for $\mathcal{U}$ and $\mathcal{V}$ rotations, respectively.
One orients a laser with Rabi frequency $\Omega_1$ in the $z$-direction and a laser with Rabi frequency $\Omega_2$ oriented in the $x$-$y$ plane.
For the second laser to address transitions distinguishable from the $\ket{g} \leftrightarrow \ket{e}$ transition while still achieving the desired momentum impulses $\pm \hbar k$, the $x$-component of this laser's wave number must satisfy $\vec{k}_2 \cdot \hat{x} \approx k$. 
Assuming both lasers have the same wavelength that addresses the $\ket{i} \leftrightarrow \ket{f}$ transition for $i \in \{ g, e \}$, the Hamiltonian in the rotating frame of the lasers is given by~\cite{Steck}
\begin{equation}
    \begin{aligned} 
\hat{H}_i = & \sum_{j = 1}^N - \hbar \Delta_i \op{f}{f}_j + \frac{\hbar}{2} \left( \Omega_1 \op{i}{f}_j + \mathrm{H.c.} \right) \\
&+ \frac{\hbar}{2} \left( \Omega_2 \hat{s}_j^+ \op{i}{f}_j + \mathrm{H.c.} \right),
    \end{aligned}
\end{equation}
where $\Delta_i$ is the detuning of the lasers from the respective transition's frequency and $\hat{s}_j^+ = \op{\hbar k / 2}{- \hbar k / 2}_j = (\hat{s}_j^-)^{\dagger}$ is the momentum shift operator.
This assumes that $\abs{\Omega_2} \ll \hbar k^2 / m$ such that the atoms remain in the $p_x = \pm \hbar k / 2$ manifold.
We take $\abs{\Delta_i} \gg \abs{\Omega_1}, \abs{\Omega_2}$ such that we can eliminate the auxiliary state to arrive at~\cite{Steck}
\begin{equation}
    \begin{aligned} 
\tilde{\hat{H}}_{i} = & \sum_{j = 1}^N \frac{\hbar}{4 \Delta_i} \left( \abs{\Omega_1}^2 + \abs{\Omega_2}^2 \hat{s}_j^+ \hat{s}_j^- \right) \op{i}{i}_j \\
&+ \frac{\hbar}{4 \Delta_i} \left( \Omega_1^* \Omega_2 \hat{s}_j^+ + \Omega_1 \Omega_2^* \hat{s}_j^- \right) \op{i}{i}_j.
    \end{aligned}
\end{equation}
The first line represents ac Stark shifts while the second line is the desired momentum exchange interaction.
Since the effective identity operator is now $\hat{\mathbb{I}}_i = \sum_j \left( \hat{s}_j^+ \hat{s}_j^- + \hat{s}_j^- \hat{s}_j^+ \right) \op{i}{i}_j$, we can shift all energies by $\hbar (\abs{\Omega_1}^2 - \abs{\Omega_2}^2 / 2) / (4 \Delta_i)$ to arrive at
\begin{equation}
    \tilde{\hat{H}}_i = \frac{\hbar}{4 \Delta_i} \sum_{j = 1}^N \left[ \frac{\abs{\Omega_2}^2}{2} \hat{s}_j^z + \left( \Omega_1^* \Omega_2 \hat{s}_j^+ + \Omega_1 \Omega_2^* \hat{s}_j^- \right) \right] \op{i}{i}_j,
\end{equation}
where we have defined $\hat{s}_j^z = \hat{s}_j^+ \hat{s}_j^- - \hat{s}_j^- \hat{s}_j^+$.
We thus have a Hamiltonian that only drives collective momentum
interactions for a desired internal level, allowing one to perform the potentially needed $\mathcal{U}$ and $\mathcal{V}$ rotations.

\section{Geometric Formalism}\label{sec:Geo}
In this section, we provide an introduction to quantum state geometry.
Quantum mechanics is often phrased algebraically, but there are important formulations in terms of Riemannian geometry~\cite{Aniello,Carinena}.
Within these formulations, the coefficients of state vectors provide coordinates for a manifold, whereupon the vector space structure of the Hilbert space is used to derive notions of curvature and distance.
In finite dimensional quantum systems with $d$ discrete states, there is an implicit map from the Hilbert space $\mh$ to the real numbers $\mathbb{R}^{2d}$ as a differential manifold. 
When considering normalized states with no global phase, $\mh$ then maps to to $\CPd$ as a manifold. 
Here, $\CPd$ is the complex projective space of $d$-dimensional c-numbers with unit norm and no global phase, i.e. $\CPd \subset \mathbb{R}^{2d}$ is the space of rays representing physical quantum states.
The vector space structure of $\mh$ provides the inner product between two states which may then be used to define tensors on $\mh$ as a manifold.
From this, one may use the coefficients of basis states $\ket{e_{\mu}}$ to expand any state as $\ket{\psi(\textbf{z})} = z^{\mu} \ket{e_{\mu}}$, with $\textbf{z} \in \CPd$ or $\mathbb{C}^d$ depending on normalization and global phase. 
Now, the inner product provides a notion of differential path length
\begin{equation}
\begin{aligned}
\quad f^2 =& | \ip{\psi(\textbf{z})}{\psi(\textbf{z}+d\textbf{z})} |^2 \\
d\zeta^2 \equiv& 1-f^2 = Q_{\mu\nu} \ dz^\mu dz^\nu \\
Q_{\mu\nu} \equiv& \bra{ \partial_\mu \psi(\textbf{z}) } \left( \hat{\mathbb{I}} - \op{\psi(\textbf{z})} \right) \ket{\partial_\nu \psi(\textbf{z})},
\end{aligned}
\end{equation}
where $Q_{\mu\nu}$ is the quantum geometric tensor~\cite{Provost,Kolodrubetz}.
The real part of $Q_{\mu \nu}$ gives the canonical Fubini-Study metric via $ds^2 \propto \Re(d\zeta^2)$, while the imaginary part provides an exact symplectic form corresponding to the Berry curvature~\cite{Kolodrubetz}.

While these formulations are powerful in their generality, dynamics are rephrased from algebraic equations of motion into $2d$ many differential equations.
Our formulation instead inherits the coordinates for $\mh$ from a Lie group with unitary representation over $\mh$.
This inheritance is equivalent to the construction using a $C^*$ algebra~\cite{Aniello}, but here we reformulate much of the construction to instead naturally fit the language of quantum metrology.
Now, the group coordinates which define unique states provide the coordinates on the the quantum state manifold, and locally return a sub-manifold of $\mathbb{R}^{2d}$. 
However, nearly all important calculations can be done on the group level.
In other words, the underlying manifold structure is inhereted through a chosen representation of the Lie group, where the group-manifold provides the underlying geometry to the Hilbert space.
Furthermore, for the case that one chooses the unitary group for which $\mh$ is the defining representation, e.g. a two dimensional Hilbert space and $\G = \mathrm{SU}(2)$, these two approaches become mechanically identical.

\begin{figure}
    \centerline{\includegraphics[width=\linewidth]{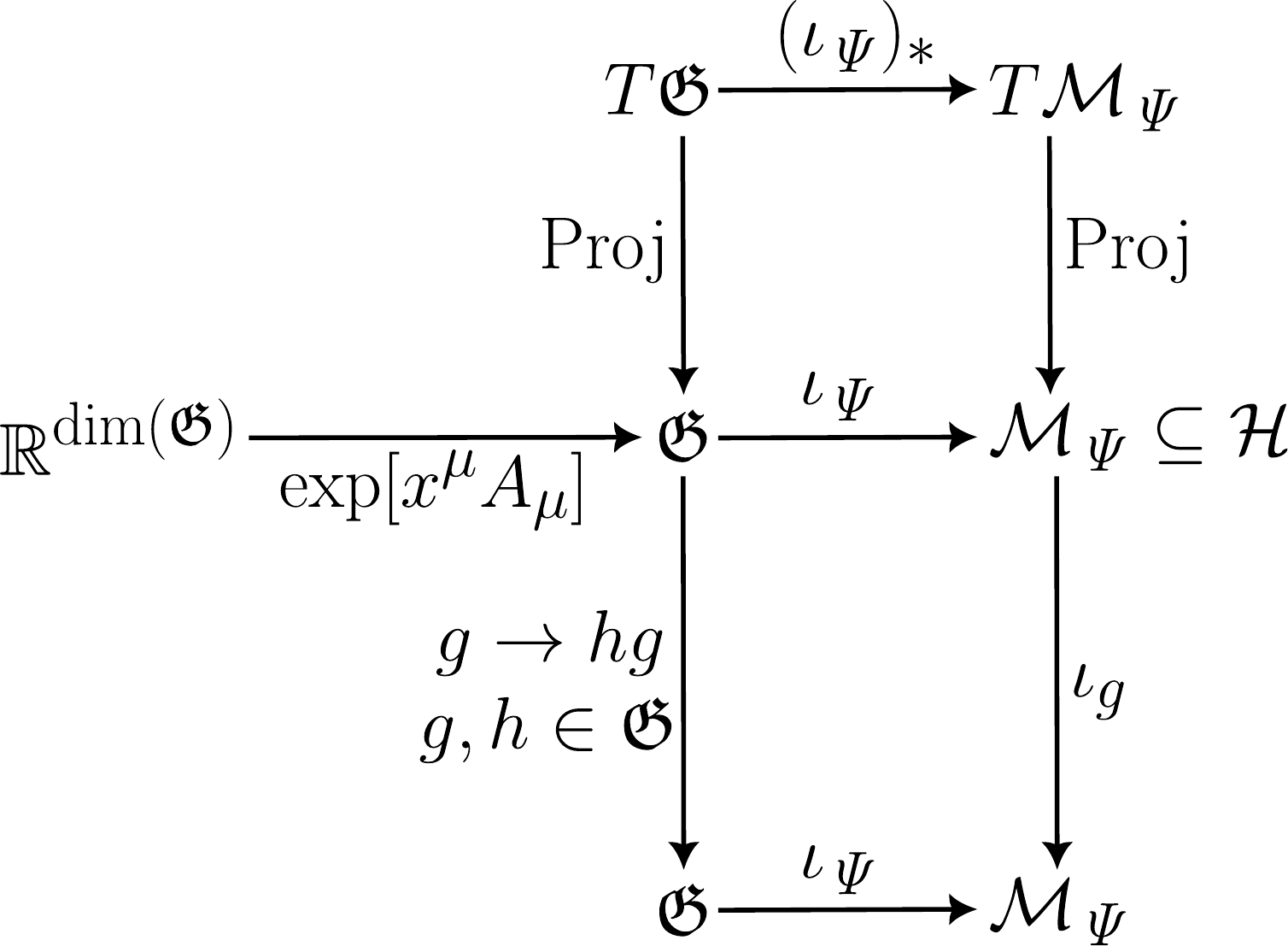}}
    \caption{
    The commutative diagram of maps between each manifold.
    $\mathrm{Proj}$ stands for projecting onto the manifold.
    } 
    \label{fig:maps}
\end{figure}

We now outline the mathematical objects we use with the goal of providing intuition for why they're useful and how they're connected.
We explore the case of a finite $d$-dimensional Hilbert space, $\mh$, and a Lie group, $\G$, that describes quantum evolution on this Hilbert space. 
The Hilbert space $\mh$ provides a representation space for $\G$, giving a unitary representation $\Pi: \G \rightarrow \text{U}(\mh)$.
Now, we identify an initial state $\ket{\Psi}$, whereupon one can act on the state using $g\in\G$ through the representation $\Pi[g]$.
In the language of quantum metrology, $\ket{\Psi}$ is referred to as the probe state or the fiducial state~\cite{Aniello}.
Since $\G$ is a manifold, each group element $g$ may be identified with coordinates $\textbf{x}\in\mathbb{R}^{\text{dim}(\G)}$, such that $g(\textbf{x})\in\G$.
Through the representation of the group action, these coordinates may be mapped into $\mh$ via $\ket{\psi(\textbf{x})} \equiv \Pi[g(\textbf{x})] \ket{\Psi}$.
Now, each state is parameterized by the coordinates of its associated group element. The set of all states realizable in this way forms our manifold $\mathcal{M}_{\Psi} \subseteq\mh$, as defined explicitly in Sec.~\ref{sec:statemfd}.
These coordinate dependencies, and the maps from one manifold to another are shown in Fig.~\ref{fig:maps}, with the map $\iota_\Psi(g) = \Pi[g] \ket{\Psi}$ being the explicit map from the group to the Hilbert space.

By doing measurements on states reached from group evolution on the probe state, one may construct probability distributions over $\mathcal{M}_\Psi$.
The precision for which one may infer the parameters $\textbf{x}$ with a given measurement basis is diagnosed by the Classical Fisher Information Matrix (CFIM).
If one further optimizes over all possible measurement bases, we recover the Quantum Fisher Information Matrix (QFIM) $\bm{\mathcal{F}}$.

Intuitively, the CFIM diagnoses how well one may distinguish probability distributions differentially far away from one another in parameter space.
The CFIM may have a non-trivial kernel, in which case measurements on $\mh$ don't resolve a corresponding coordinate change in $\G$.
This is the case when one has phases which may not be measured by in the chosen measurement basis, or to a global phase which may never be measured.
By removing the kernel of the CFIM, one recovers a well-defined metric on these probability distributions over $\mathcal{M}_\Psi$.

The QFIM, similarly, defines how well one could optimally distinguish between two states which are differentially far away in parameter space, and thus may be made into to a Riemannian metric on these states.
Similar to the CFIM, some parameters may correspond to the kernel of the QFIM, such as when a Lie group element $g$ acts on a state which is an eigenvector of the corresponding Lie algebra member: $g(\textbf{x}) \ket{\psi} = e^{-i \theta(\textbf{x})} \ket{\psi}$ for global phase $\theta(\textbf{x})$.
Once again the removal of these elements returns a well-defined metric which now lives on the manifold $\mathcal{M}_{\Psi}$.

The group coordinates which correspond to unique states on $\mathcal{M}_\Psi$ become coordinates on the quantum state manifold.
For two coordinates $x^\mu$ and $x^\nu$ of $\textbf{x} = (x^0, ...,x^\mu,...,x^\nu,...)$, this optimization yields
\begin{equation} \label{eq:QFIM}
    \mathcal{F}_{\mu \nu} = \frac{1}{2} \Tr[ \hat{\rho} \{ \hat{L}_\mu, \hat{L}_\nu \} ], \quad \partial_\mu \hat{\rho} = \frac{1}{2} \left( \hat{\rho} \hat{L}_\mu + \hat{L}_\mu \hat{\rho} \right),
\end{equation}
where $\{\hat{A},\hat{B}\} = \hat{A} \hat{B} + \hat{B} \hat{A}$ is the anti-commutator and $\hat{L}_\mu$ is the Symmetric Logarithmic Derivative (SLD)~\cite{Braunstein} with respect to the coordinate $x^\mu$.
Geometrically, $\mathcal{F}_{\mu \nu}$ provides a metric tensor over the space of density matrices,
\begin{equation}
    ds^2 = \mathcal{F}_{\mu\nu} dx^\mu dx^\nu,
\end{equation}
and is equivalent to four times the Bures metric for mixed states.
For pure states, one finds that $\hat{L}_\mu = 2 \partial_\mu \hat{\rho}$, and so~\cite{Facchi}
\begin{equation}
    \mathcal{F}_{\mu \nu} = 4 \Re( Q_{\mu\nu} ),
\end{equation}
matching the Fubini-Study metric (up to rescaling).

\subsection{The Lie Group}
We now outline the relevant Lie group structures.
In general, any quantum evolution may be described by the appropriate Lie group represented unitarily on the Hilbert space.
Let $\G$ be some Lie group with $g\in\G$ being some group element.
For applications to quantum metrology, it suffices to only consider simply connected Lie groups.
We may choose coordinates on $\G$ by specifying a basis for the Lie Algebra of $\G$ given by the vector space $\mathfrak{g}$.
We choose basis vectors $\{A_\mu\}\subset\mathfrak{g}$ such that $\text{span}(\{A_\mu\}) = \mathfrak{g}$.
For concreteness, if $\G = \text{SU}(n)$ then one could construct, for example, the basis operators $A_\mu$ according to Sec.~\ref{OperatorBasis}.
Now, any group element $g\in\G$ can be identified via
\begin{equation}
    g = \exp(x^\mu A_\mu),
\end{equation}
with real numbers $x^\mu$. 
From this, it is clear that any change of basis on $\mathfrak{g}$ has a corresponding change of coordinates.
For example, if $\{B_\nu\} \subset \mathfrak{g}$ with $\text{span}(\{B_\nu\}) = \text{span}(\{A_\mu\})$ we can find $\Lambda^\mu_\nu$ such that $B_\nu = \Lambda^\mu_\nu A_\mu$ implies $x^\mu = \Lambda^\mu_\nu y^\nu$, and thus the group element $g$ stays coordinate independent:
\begin{equation}
g = \exp(x^\mu A_\mu) = \exp(y^\nu B_\nu).
\end{equation}

Lastly, we point out that the Lie Group $\G$ carries with it a canonical metric in the form of the Killing form, defined on the Lie algebra as:
\begin{equation}
    \kappa: \mathfrak{g}\times \mathfrak{g} \rightarrow \mathbb{R}, \quad \kappa(A_\mu,A_\nu) \equiv \Tr_\mathfrak{g}[ \mathrm{ad}_{A_\mu} \circ \mathrm{ad}_{A_\nu}],
\end{equation}
for arbitrary basis vectors $A_\mu,A_\nu\in\mathfrak{g}$ and with the trace taken over the Lie algebra as a vector space. The function $\mathrm{ad}_{A_\mu}$ is the adjoint action of $A_\mu$ on the Lie algebra, which is the usual commutator $\mathrm{ad}_{A_\mu}(A_\nu) \equiv [A_\mu,A_\nu]$.
The general properties of $\kappa$ are described in Ref.~\cite{Gorbatsevich}, but notably for $\mathrm{SU}(n)$, one has $\kappa(A_\mu,A_\nu) \equiv 2n \Tr[A_\mu A_\nu]$ with the trace being taken over the vector space of the defining representation.
Thus, $\kappa$ provides a metric on $\G$ near the identity via
\begin{equation}
(ds^2)_\G \equiv \kappa_{\mu\nu} dx^\mu dx^\nu,
\end{equation}
with $\kappa_{\mu\nu} = \kappa(A_\mu,A_\nu)$. 
For $\G = \mathrm{SU}(n)$, one has
\begin{equation}
(ds^2)_{\mathrm{SU}(n)} \equiv 2n \Tr[A_\mu A_\nu] dx^\mu dx^\nu.
\end{equation}

\subsection{The Hilbert Space} \label{sec:statemfd}
Now, we show how the structure of the Lie group is inherited to the Hilbert space through a unitary representation.
We consider a Hilbert space $\mh$ of dimension $|\mh| = d$.
Since $\mh$ is a vector space, we may choose representation map,
\begin{equation}
    \Pi:\G \rightarrow \text{U}(\mh),
\end{equation}
which implicitly carries with it the representation of $\mathfrak{g}$: $R:\mathfrak{g} \rightarrow \mathfrak{u}(\mh)$.
The representation $R$ acts on our basis vectors $A_\mu$ according to
\begin{equation}
    R(A_\mu) = -i \hat{G}_\mu,
\end{equation}
where $\hat{G}_\mu$ is a Hermitian operator on $\mh$. 
Here, we note that this factor of $-i$ accounts for the difference in convention between physics and mathematics and is representation independent, but the relation from $A_\mu$ to $\hat{G}_\mu$ is notably representation dependent. 
The exact relationship between each of these representation maps is
\begin{equation}
\begin{aligned}
\Pi[g] =& \Pi[ \exp(x^\mu A_\mu) ]  = e^{x^\mu R(A_\mu) } = e^{-i x^\mu \hat{G}_\mu } .
\end{aligned}
\end{equation}
The choice of basis vectors on $\mathfrak{g}$ is thus equivalent to a choice of basis for Hermitian matrices on $\mh$.

To parameterize $\mh$, we first choose a starting state, $\ket{\Psi} \in \mh$, and define the map \begin{equation}
\begin{aligned}
     & \iota: \mh\times \G \rightarrow \mh, \\
    \text{where} \quad & \iota_\Psi: \G \rightarrow \mh \quad \text{and} \quad \iota_g: \mh \rightarrow \mh .
\end{aligned}
\end{equation} 
Now, $\iota_\Psi$ is a map between manifolds from which we can push forward vectors or pull back differentials, while $\iota_g$ is a map from $\mathcal{H}$ to itself, which allows relevant objects to be pushed forward and pulled back to $\Psi$.
This definition matches the definition of $\iota_Q$ in Ref.~\cite[Eq.~(2.7)]{Aniello}.
The map $\iota_\Psi$ takes $g$ as an argument and is defined as the action of $\Pi[g]$ on $\ket{\Psi}\in\mh$:
\begin{equation}
    \iota_\Psi(g) \equiv \Pi (g) \ket{\Psi} = e^{-i x^\mu \hat{G}_\mu } \ket{\Psi}.
\end{equation}
Meanwhile, $\iota_g$ takes any arbitrary state $\psi$ as an argument and is similarly defined as the action of $\Pi[g]$ on $\ket{\psi}$:
\begin{equation}
    \iota_g (\psi) \equiv \Pi (g) \ket{\psi}.
\end{equation}
Clearly, $\iota(\psi,g) = \iota_\psi (g) = \iota_g (\psi)$, but both $\iota_\psi (g)$ and $\iota_g (\psi)$ are needed for properly defined push forward and pull backs of tensors.

This map now allows us to define the manifold on which one wishes to carry out quantum metrology,
\begin{equation}
    \mathcal{M}_{\Psi} = \{ \ket{\psi_g} \ \text{s.t.} \ket{\psi_g} = \iota_\Psi(g) \}.
\end{equation}
This manifold $\mathcal{M}_{\Psi} \subseteq \mh$ is the group orbits of $\ket{\Psi}$ under all possible group actions in $\mathfrak{G}$; i.e., it is the set of states that may be reached under group actions when $\ket{\Psi}$ is the probe state.

Within $\G$, there is a subgroup which will act trivially on any given state by only adding a global phase.
Following Ref.~\cite{Provost}, the Lie group can be divided into this subgroup and the rest of it's actions.
The subgroup which acts trivially is the stabilizer group of the state, and this stabilizing group is generated by a sub-algebra for which the has the quantum state is an eigenvector. 
The group elements which induce non-trivial evolution are generated by the remaining members of $\mathfrak{g}$.
By removing the members of the sub-algebra generating the stabilizing group, and their corresponding coordinates, one is left with well defined coordinates on $\mathcal{M}_\Psi$. 
Without this, the group coordinates only parameterize the states and do not become well defined coordinates.

Now, by considering one parameter paths through $\G$ we can use $\iota_\Psi$ to identify paths in $\mathcal{M}_{\Psi}$.
Let 
\begin{equation}
\gamma:\mathbb{R}\rightarrow \G, \quad \gamma(t) = \exp(x^\mu(t) A_\mu),
\end{equation}
so that the composition of $\iota_\Psi$ with $\gamma$ yields $\iota_\Psi\circ\gamma: \mathbb{R} \rightarrow \mathcal{M}_\Psi$. 
Explicitly,
\begin{equation}
(\iota_\Psi\circ\gamma)(t) = \ket{\psi(t)} \equiv e^{-i x^{\mu}(t) \hat{G}_\mu} \ket{\Psi}
\end{equation}
and if our path has $\iota_\Psi\circ\gamma(0) = \ket{\Psi}$, then
\begin{equation}\label{eq:schrod}
\begin{aligned}
\frac{\partial}{\partial t} \ket{\psi(t)} & = \frac{\partial x^\mu(t) }{\partial t} \partial_\mu \ket{\psi(t)} \\
& = \frac{\partial x^\mu(t) }{\partial t} (-i \hat{G}_\mu) \ket{\psi(t)}.
\end{aligned}
\end{equation}
Now, derivatives of $\iota_\Psi\circ\gamma$ for any possible path, $\gamma$, defines all possible tangent vectors to the point $\Psi$.
To find tangent vectors at any other point in $\mathcal{M}_\Psi$, one may use $\iota_g$ to push forward the tangent vectors at $\ket{\Psi}$.
In the case that $x^\mu = \omega^\mu t$ where $\hbar \omega^\mu$ is energies corresponding to the operator $\hat{G}_\mu$ in the Hamiltonian, then $\frac{\partial x^\mu }{\partial t} = \omega^\mu$ and Eq.~\eqref{eq:schrod} would correspond to the familiar case of a time-independent Hamiltonian.

Furthermore, from the equation
\begin{equation}
\frac{\partial x^\mu }{\partial t} \partial_\mu e^{-i x^\mu(t) \hat{G}_\mu } \ket{\Psi} = \frac{\partial x^\mu }{\partial t} \left( -i \hat{G}_\mu \right) e^{-i x^\mu(t) \hat{G}_\mu } \ket{\Psi},
\end{equation}
we can match terms and use $\gamma(t)$ to define paths through any point, who's derivatives define a basis for tangent vectors at any point. This yields a basis for vectors at each point on $\mathcal{M}_\Psi$, and vector fields are now given by the standard identification
\begin{equation}
    \partial_\mu \equiv -i \hat{G}_\mu.
\end{equation}

These tangent vector fields could have equivalently been found via push forward.
Any vector $\vec{V} = V^\mu \partial_\mu \in \mathfrak{g}$ under $(\iota_\Psi)_*$:
\begin{equation}
    (\iota_\Psi)_* [ \vec{V} ] = V^\mu (\iota_\Psi)_* [ \partial_\mu ] = -i V^\mu \hat{G}_\mu \ket{\Psi},
\end{equation}
where $(\iota_\Psi)_* [ \vec{V} ] \in T_\Psi \mathcal{M}_\Psi$, with $T_\Psi \mathcal{M}_\Psi$ being the tangent vector space to the point $\Psi$ on the manifold $\mathcal{M}_\Psi$.
We can pushfoward the tangent vector fields $T \G$ to tangent vector fields on the state manifold $T \mathcal{M}_{\Psi}$, as shown in Fig.~\ref{fig:maps}.

\subsection{The Fisher Information Metric}
Lastly, we show the interplay of the Lie group's geometry and the Fisher Information Metric.
The QFIM has been shown to be a unique Riemannian metric on the space of pure quantum states and a well defined metric on the space of mixed states.
We will use pure states to show the pull back of the metric to the Lie group, but thus far all the maps defined would work suitably well for unitary evolution of an initially mixed state.
The case of evolution with decoherent dynamics is left for future work.

We now derive the equations in the Main Text and show that the normalization of vectors on $\mh$ is a naturally determined via the Killing Form on $\G$.
Specifically, we use the fact that a metric serves as the tensorial promotion of the inner product,
\begin{equation} \label{InnerProduct}
    \begin{aligned}
\langle \vec{V}, \vec{W} \rangle_{\mathbf{x}} &= \langle \partial_\mu, \partial_\nu \rangle_{\mathbf{x}} V^\mu W^\nu \\
&= \mathcal{F}_{\mu\nu} V^\mu W^\nu,
    \end{aligned}
\end{equation}
where now, for pure states, $\mathcal{F}_{\mu\nu}$ is given by
\begin{equation}
    \begin{aligned}
\mathcal{F}_{\mu \nu} &= 4 \,\mathrm{cov}_{\psi_g}( \hat{G}_\mu, \hat{G}_\nu) \\
&= 2 \exv{ \{ \hat{G}_\mu , \hat{G}_\nu \} }_{\psi_g} - 4 \exv{ \hat{G}_\mu }_{\psi_g} \exv{ \hat{G}_\nu }_{\psi_g}.
    \end{aligned}
\end{equation}
This formula may be equivalently derived from the fact that derivatives are pushed forward from $\G$, or equivalently, the Lie Algebra is represented as operators on $\mh$.

The tensor $\mathcal{F}_{\mu\nu}$ does not yet behave as a matrix, and therefore one needs to raise an index.
Formally, one may use the pull back $(\iota_\Psi)^*$ to map $\mathcal{F}_{\mu\nu}$ to a tensor on $\G$, as shown schematically for covector spaces in Fig.~\ref{fig:maps}(c).
By pulling back the metric from $\mathcal{M}_\Psi$ to $\G$ as a tensor using $(\iota_{\Psi})^*$, one can use the metric defined by $\kappa_{\mu\nu}$ to raise one index on $\mathcal{F}_{\mu\nu}$.
Using $(\kappa^{-1})^{\alpha\beta}$, which is defined implicitly by $(\kappa^{-1})^{\mu\alpha}\kappa_{\alpha\nu}=\delta^\mu_\nu$, one finds
\begin{equation}
    \mathcal{F}^\mu_\nu = (\kappa^{-1})^{\mu \alpha} \ (\iota_\Psi)^*[ \mathcal{F}_{\alpha\nu} ].
\end{equation}
This is a (1,1) tensor which now has well defined eigenvectors and eigenvalues.
Diagonalizing this tensor is equivalent to diagonalizing the QFIM, $\bm{\mathcal{F}}$, but now with proper dependence on the inner product on $\mathfrak{g}$.
This imposes the normalization properties used in Sec.~\ref{OperatorBasis}.

Since $\mathcal{F}_{\mu\nu}$ defines a metric over pure states in $\mh$, one has that the vector whose inner product is maximized under this inner product is naturally the direction from $\G$ which the state will evolve the fastest.
By diagonalizing the tensor $\mathcal{F}^\mu_\nu$ on $\G$, one finds the vector whose inner product is maximized on $\mh$ by equivalently finding the largest eigenvalue of $\bm{\mathcal{F}}$ over the entire Lie Algebra.
Here, the basis of the Lie Algebra must be orthonormal with respect to the inner product derived from $\kappa$; otherwise, scaling in the QFI could be due to an arbitrary rescaling of operators.
Using the construction given in Sec.~\ref{OperatorBasis}, one finds that this normalization for $\mathrm{SU} (n)$ is given by
\begin{equation} \label{eq:KillingNorm}
    \Tr[\hat{G}_\mu \hat{G}_\nu] = \delta_{\mu\nu} \frac{1}{(n + 1)!} \prod_{j = 1}^n (N + j),
\end{equation}
which is equivalent to requiring that the basis elements of the Lie Algebra are each members of an $N+1$ dimensional representation of $\mathfrak{su}(2) \subset \mathfrak{su}(n)$.
Physically, this normalization guaruntees that subsystems of the Hilbert space which form qubits, or which represent symmetric collections of qubits, accumulate relative phase at unit rates.
This fixes the standard quantum limit scaling for unentangled generalized coherent spin states at $N$, and the Heisenberg limit scaling for GHZ-like states at $N^2$. 
This normalization additionally guarantees that the QFI may act as an entanglement witness, with the sufficient condition being that any eigenvalue of $\bm{\mathcal{F}}$, $\lambda_i$, scales as $\lambda_i>N$.

In summary, vectors come from $\G$ where they are plugged into a state-dependent inner product as vectors on $\mh$ via the push forward.
These vectors define all the directions in which a state may evolve.
Many of these directions will be null-directions in $\mathcal{M}_\Psi$ and may therefore be ignored, leaving only well defined coordinate directions on $\mathcal{M}_\Psi$.
Of the vectors which are non-zero, many will be non-orthogonal under $\mathcal{F}_{\mu\nu}$'s inner product Eq.~\eqref{InnerProduct}.
This means that those directions of evolution are partially degenerate according to how they evolve the state.
By pulling back $\mathcal{F}_{\mu\nu}$ to $\G$ and using the Killing form to treat $\mathcal{F}_{\mu\nu}$ as a matrix, $\bm{\mathcal{F}}$, one may diagonalize the QFIM.
By diagonalizing $\bm{\mathcal{F}}$, one finds the eigenvalues representing the QFI.
In doing so, one is left with a matrix whose eigenvectors are orthogonal vectors according to the Information metric's inner product, and whose eigenvalues are the distance, $ds$, which one will evolve in for a differential change in the corresponding coordinate.
These vectors are normalized with respect to the Killing form's inner product, which for $\mathrm{SU}(n)$ is the trace inner product of operators.
This guarantees that improvements in scaling of the QFI are truly a result of entanglement and not an arbitrary rescaling.
Thus, a small change in $x^\mu$ corresponds to a large pathlength $ds$ over our Hilbert space, which can be well distinguished by quantum measurement! Finding the coordinates for which $\mathcal{F}_{\mu\nu}$ is maximum and diagonal corresponds to the eigenvalues of the QFIM.

\bibliography{references.bib}